\documentclass{elsart}
\usepackage{graphicx}
\usepackage{dcolumn}
\usepackage{bm}
\begin{document}
\begin{frontmatter}
\title{Search for neutrinoless decays\\ $\tau\to \ell h h$ and $\tau\to \ell V^0$} 
\collab{Belle Collaboration}
  \author[VPI]{Y.~Yusa}, 
  \author[KEK]{K.~Abe}, 
  \author[TohokuGakuin]{K.~Abe}, 
  \author[KEK]{I.~Adachi}, 
  \author[Tokyo]{H.~Aihara}, 
  \author[Tsukuba]{Y.~Asano}, 
  \author[BINP]{V.~Aulchenko}, 
  \author[Lausanne]{A.~Bay}, 
  \author[JSI]{U.~Bitenc}, 
  \author[JSI]{I.~Bizjak}, 
% ...

\address[BINP]{Budker Institute of Nuclear Physics, Novosibirsk, Russia}
% ...
\address[KEK]{High Energy Accelerator Research Organization (KEK), Tsukuba, Japan}
% ...
\address[JSI]{J. Stefan Institute, Ljubljana, Slovenia}
% ...
\address[Lausanne]{Swiss Federal Institute of Technology of Lausanne, EPFL, Lausanne, Switzerland}
% ...
\address[TohokuGakuin]{Tohoku Gakuin University, Tagajo, Japan}
% ...
\address[Tokyo]{Department of Physics, University of Tokyo, Tokyo, Japan}
% ...
\address[Tsukuba]{University of Tsukuba, Tsukuba, Japan}
\address[VPI]{Virginia Polytechnic Institute and State University, Blacksburg, VA, USA}
% ...

\begin{abstract}
We have searched for neutrinoless $\tau$ lepton decays into 
$\ell h h$ or $\ell V^0$, where $\ell$ stands for an 
electron or muon, $h$ for a charged light hadron, $\pi$ or $K$, 
and $V^0$ for a neutral vector meson, $\rho^0$, $K^*(892)^0$ and $\phi$, 
using a 158 fb$^{-1}$ data sample collected with the 
Belle detector at the KEKB  $e^+e^-$ collider. Since the number of 
events observed are consistent with the expected background, 
we set upper limits on the branching fractions in the range of 
$(1.6-8.0) \times 10^{-7}$ for various decay modes at the 90\% confidence level.
\end{abstract}

\end{frontmatter}

\section{Introduction}
In the Standard Model (SM), lepton-flavor-violating (LFV) 
decays of charged leptons are forbidden, or highly suppressed even if 
the effect of neutrino mixing is taken into account~\cite{SMLFV}. In contrast, LFV 
decay processes are expected to appear with much larger branching fractions than those in the SM
if there are contributions from new physics. Searches for LFV decay processes may thus reveal new 
physics beyond the SM. Some models predict LFV decays of  $\tau$ leptons at a level accessible 
at the high luminosity $B$-factories \cite{seesawilakovac96,seesawilakovac00}. 
In this paper, we report on a search for LFV in fourteen $\tau^-$ decay modes into 
neutrinoless final states with one charged lepton $\ell$ and two charged pseudoscalar mesons $h$:  
$e^- \pi^+ \pi^-$, $e^+ \pi^- \pi^-$, $\mu^- \pi^+ \pi^-$, $\mu^+ \pi^- \pi^-$, $e^- \pi^+ K^-$, 
$e^- \pi^- K^+$, $e^+ \pi^- K^-$, $e^- K^+ K^-$, $e^+ K^- K^-$, $\mu^- \pi^+ K^-$, $\mu^- \pi^- K^+$, 
$\mu^+ \pi^- K^-$, $\mu^- K^+ K^-$ and $\mu^+ K^- K^-$, and  eight modes in which $\tau$ decays 
into one lepton and one vector meson: $e^- \rho^0$, $e^- K^*(892)^0$, $e^- \bar{K}^*(892)^0$, $e^- \phi$, 
$\mu^- \rho^0$, $\mu^- K^*(892)^0$, $\mu^- \bar{K}^*(892)^0$ 
and $\mu^- \phi$.\footnote{Charge conjugate decay modes are implied throughout the paper.} 
Current upper bounds on the branching fractions for these decays 
are of the order of $10^{-6}$ at 90\% confidence level (CL) 
and have been set in the CLEO experiment using a data sample of 
4.79 fb$^{-1}$~\cite{CLEOLHH}. 
Very recently CLEO results on $\tau \to \ell h h $ modes were improved by the BaBar experiment 
and upper limits in the range $(0.7-4.8) \times 10^{-7}$ were obtained from a 
$221.4~{\rm fb}^{-1}$ data sample \cite{BABARLHH}. 
We present here results of a new search based on a data sample of 158.0 fb$^{-1}$ corresponding to
$140.9 \times 10^6$ $\tau$-pairs collected with the Belle detector~\cite{BELLE} 
at the KEKB asymmetric-energy $e^{+}e^{-}$ collider~\cite{KEKB} 
operating at or near the $\Upsilon(4S)$ resonance. 
\section{Event Selection}
 The Belle detector is a general purpose detector with excellent 
capabilities for precise vertex determination and particle 
identification. Tracking of charged particles is performed using a 
three-layer double-sided silicon vertex detector (SVD) and a 
fifty-layer cylindrical drift chamber (CDC) located in a 1.5 T 
magnetic field. Charged hadrons are identified by combining 
$dE/dx$ information from the CDC, signal pulse-heights from 
aerogel $\check{\rm C}$erenkov counters (ACC) and timing 
information from time-of-flight 
scintillation counters (TOF). Photons are 
reconstructed using a CsI(Tl) electromagnetic calorimeter (ECL). Muons are 
detected by fourteen layers of resistive plate counters interleaved 
with iron plates (KLM). 

We use TAUOLA \cite{KORALB} for Monte Carlo (MC) event generation 
of $\tau$-pair signals and KKMC \cite{KKMC} to implement initial and final state radiation.
The KKMC MC program predicts a cross-section of 
$\sigma (e^+e^- \to \tau^+ \tau^-)= (0.8916 \pm 0.0006)~{\rm nb}$ at the center-of-mass energy of KEKB. 
We calculate the number of $\tau$-pair events from the cross section and measured integrated luminosity 
$\mathcal{L}_{\rm int} = 158.01$ fb$^{-1}$.
The MC data is processed through the Belle detector simulation program based on GEANT3 \cite{GEANT} 
to determine signal efficiencies.  We use the CLEO QQ event generator \cite{QQ} for 
hadronic events and AAFHB \cite{aafhb} for two-photon events, and study their 
contributions to the background of each $\tau$ decay mode. 

We search for $\tau$-pair events in which one $\tau$ decays into the 
$\ell h h$ (3-prong) final state. The other $\tau$ dominantly decays into one charged particle
and any number of neutrals (1-prong) with a branching fraction of 
85.35\% \cite{PDG}. We require that there be four charged tracks with 
zero net charge and any number of photons in an event.
We reconstruct the trajectory of a charged track from hits in the 
SVD and CDC, and require that a reconstructed transverse momentum 
be larger than 0.1 GeV/$c$ and polar angle $\theta$ be 
within the range 25$^\circ < \theta < 140^\circ$, with respect to the direction opposite to the 
$e^+$ beam. For all charged tracks, the distance of the closest approach to 
the interaction point (IP) is required to be within 1 cm transversely 
and 3 cm along the $e^+$ beam. 
Photons are selected from neutral ECL clusters with an energy threshold 
$E_{\rm cluster} >$ 0.1 GeV and are separated by at least 30 cm from the extrapolated 
projection point of any charged track. 
The tracks and photons in an event are divided into two hemispheres 
in the $e^+e^-$ center-of-mass system (CMS), with a plane 
perpendicular to the thrust axis calculated from the momenta of 
all charged tracks and photons. We select 3-prong vs. 1-prong 
topology events, i.e. three charged tracks are in one
hemisphere and one charged track in the other. We define the former 
hemisphere as the signal side and the latter as the tag side. The 
number of photons on the signal side should be less than or equal to two, to allow for photons from 
initial and final state radiation or photons radiated from electron tracks.

Electrons are identified by an electron likelihood that includes the 
$dE/dx$ value measured with the CDC, the ratio of the cluster energy 
from the ECL to the track momentum measured with CDC, ACC hits and shower shape in ECL \cite{eid-NIM}. 
With this likelihood, electrons are selected with an average efficiency of 85\% over the  
whole momentum range of those in the signal MC. 
The momentum of a charged track in the laboratory system $p$ should be greater than 0.3 GeV/$c$ 
to avoid poorly identified electrons. In order to correct for the energy loss due to bremsstrahlung, 
the momentum of an electron track is recalculated adding the momentum of 
radiated photon clusters if an ECL cluster with energy less 
than 1.0 GeV is detected within a cone angle of 10$^\circ$ along the electron flight direction.

The muon likelihood is formed from two variables: the difference between the range calculated from 
the momentum of the particle and that measured with the KLM, as well as the $\chi^2$ value of the 
KLM hits with respect to the extrapolated track~\cite{muid-NIM}. The average efficiency 
is evaluated as 90\% from signal MC for muon in the momentum range for LFV decays.
For muons, $p$ should be larger than 0.6 GeV/$c$ for the same reason as in the electron case.

Tracks that do not satisfy the requirements for electron or muon candidates 
are classified as hadrons. To distinguish kaons from pions, we 
use a likelihood ratio which is calculated from $dE \slash dx $, time-of-flight and the hits in the ACC. 
We achieve kaon efficiencies of 85\% in the barrel and 80\% in the endcap region. 
In these cases, $p$ should be larger than 0.5 GeV/$c$. The remaining tracks are treated as pions. 

Vector mesons are reconstructed in the following decay modes: 
$\rho^0 \to \pi^+\pi^-$, $K^*(892)^0 \to K^+\pi^-$ and $\phi \to K^+K^-$. 
We calculate the vector meson mass, $M_V$. We then fit the $M_V$ distribution of the signal MC with 
two Gaussian distributions to take into account the effects of the intrinsic width of the 
resonances and the detector resolution. 

The width of the signal windows for each decay modes is $\pm 1.64 \sigma$, where $\sigma$ is 
the standard deviation of the broader Gaussian component: 
445 MeV/$c^2 < M_{\rho} <$ 1092 MeV/$c^2$, 
730 MeV/$c^2 < M_{K^*} <$ 1064 MeV/$c^2$ and 
1005 MeV/$c^2 < M_{\phi} < 1035$ MeV/$c^2$.

The background that remains after applying all the selection criteria based on event topology 
and particle identification is dominated by Bhabha events. 
This background is effectively reduced by requiring the invariant masses 
calculated for all combinations of two oppositely charged tracks to 
be greater than 0.2 GeV$/c^{2}$, assuming the electron mass. 

Backgrounds from $B\bar{B}$ and $c\bar{c}$ events are suppressed by the 
selection criteria listed above, however 
the background from $e^+ e^- \to q \bar{q}$ continuum events with the light quarks 
$u$, $d$ and $s$ ($uds$ continuum) is not negligible.
Since these backgrounds often include $\pi^0$ decays, we require that the number of photons 
on the tag side $n_\gamma$ not exceed 1. If the tag is an electron or muon, the $n_\gamma$ requirement 
effectively suppresses the $uds$ continuum background. 
For tags where the tag side is a pion, the selected events are investigated further as follows.

\begin{figure}
\begin{center}
\includegraphics[scale=0.42,clip]{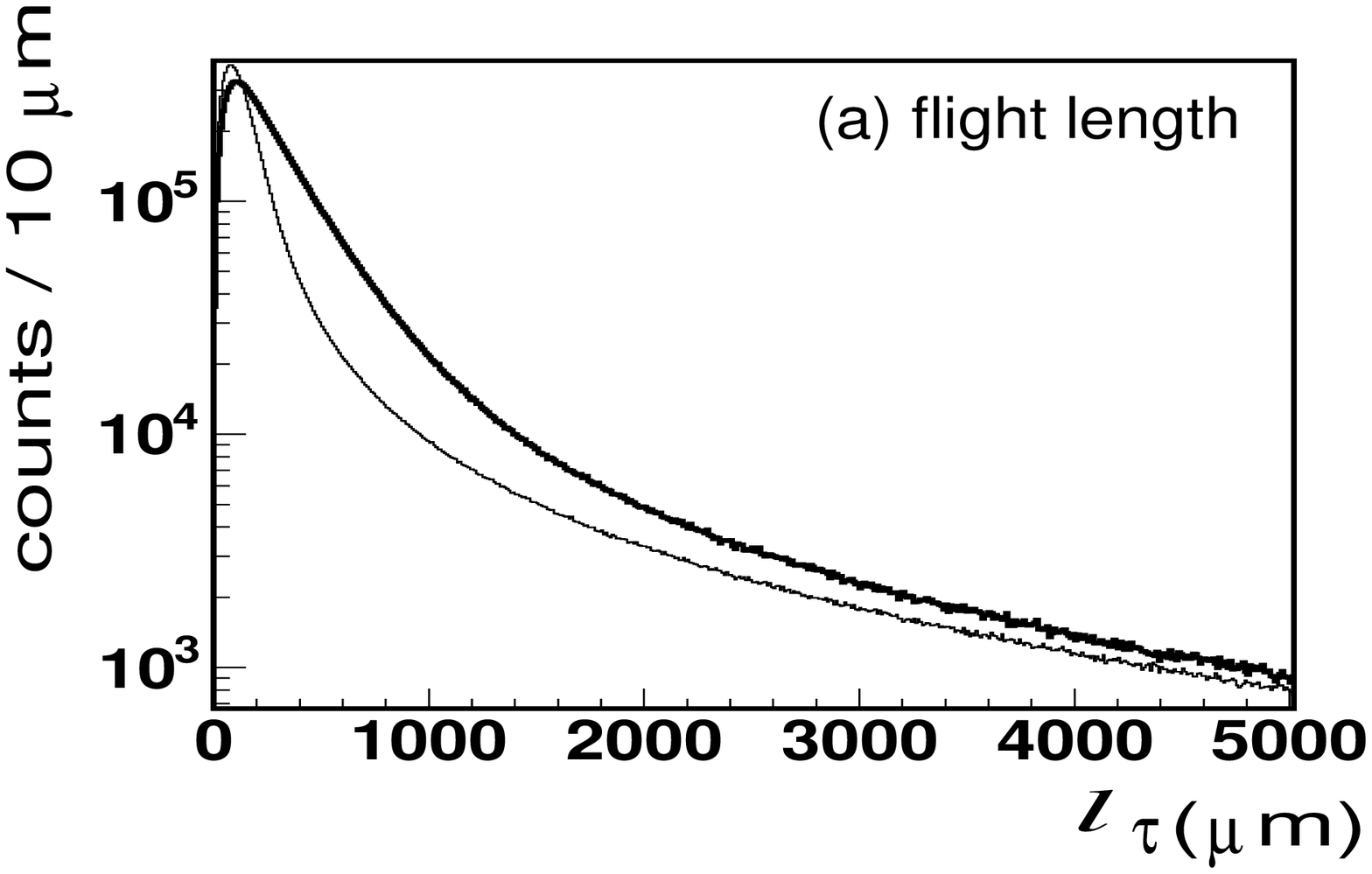}
\includegraphics[scale=0.42,clip]{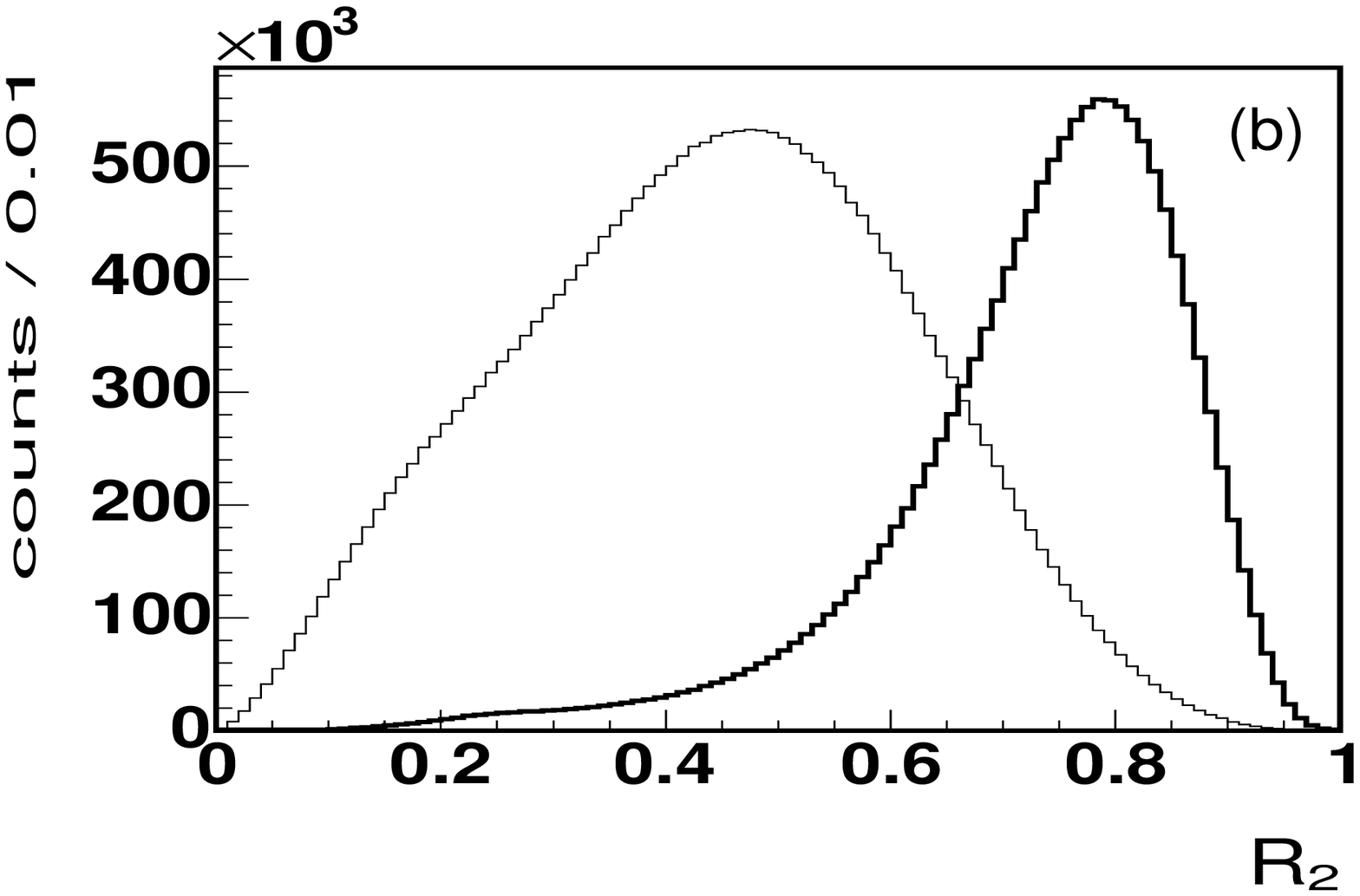}
\caption{(a) Flight length $l_\tau$ and (b) $R_2$ distributions of the signal MC events (thick line) and $uds$ continuum events (thin line). The histogram normalization is arbitrary.}
\label{fig_qqveto:flvsR2}
\end{center}
\end{figure}

We measure the $\tau$ flight length $l_{\tau}$ using information on IP 
and $\tau$ vertex position that is reconstructed from the tracks on the signal side.
Since $\tau$ leptons travel significant distance before decaying ($c\tau = 87 \mu$m), 
this feature can be used to suppress background. 
Since $\tau$-pair events have a more jet-like shape than $uds$ continuum events 
and are distinguishable by use of shape variables. 
We use a normalized second Fox-Wolfram moment $R_2$ \cite{R2} to represent 
the event shape. Figure \ref{fig_qqveto:flvsR2} shows
the $l_\tau$ and $R_2$ distributions for $\tau$-pair and $uds$ continuum 
events. We calculate the two-dimensional probability density function (PDF) in the  
$l_\tau$ vs. $R_2$ plane, and form a likelihood ratio 
$\mathcal{L}_{\tau\tau}/(\mathcal{L}_{\tau\tau} + \mathcal{L}_{uds})$, 
where $\mathcal{L}_{\tau\tau}$ and $\mathcal{L}_{uds}$ are the PDFs for $\tau$-pair 
signal and $uds$ continuum background, respectively.
The condition $\mathcal{L}_{\tau\tau}/(\mathcal{L}_{\tau\tau} + \mathcal{L}_{uds}) > 0.45$ is 
optimized using MC. This condition removes 60\% of $uds$ continuum background, while 90\% of 
the signal is kept when the track on the tag side is a pion.
After combination of all selections including the $n_\gamma$ and likelihood ratio requirements, 
the $uds$ continuum background is suppressed by a factor of $10^{5}$. 

\begin{figure}
\begin{center}
\includegraphics[scale=0.43,clip]{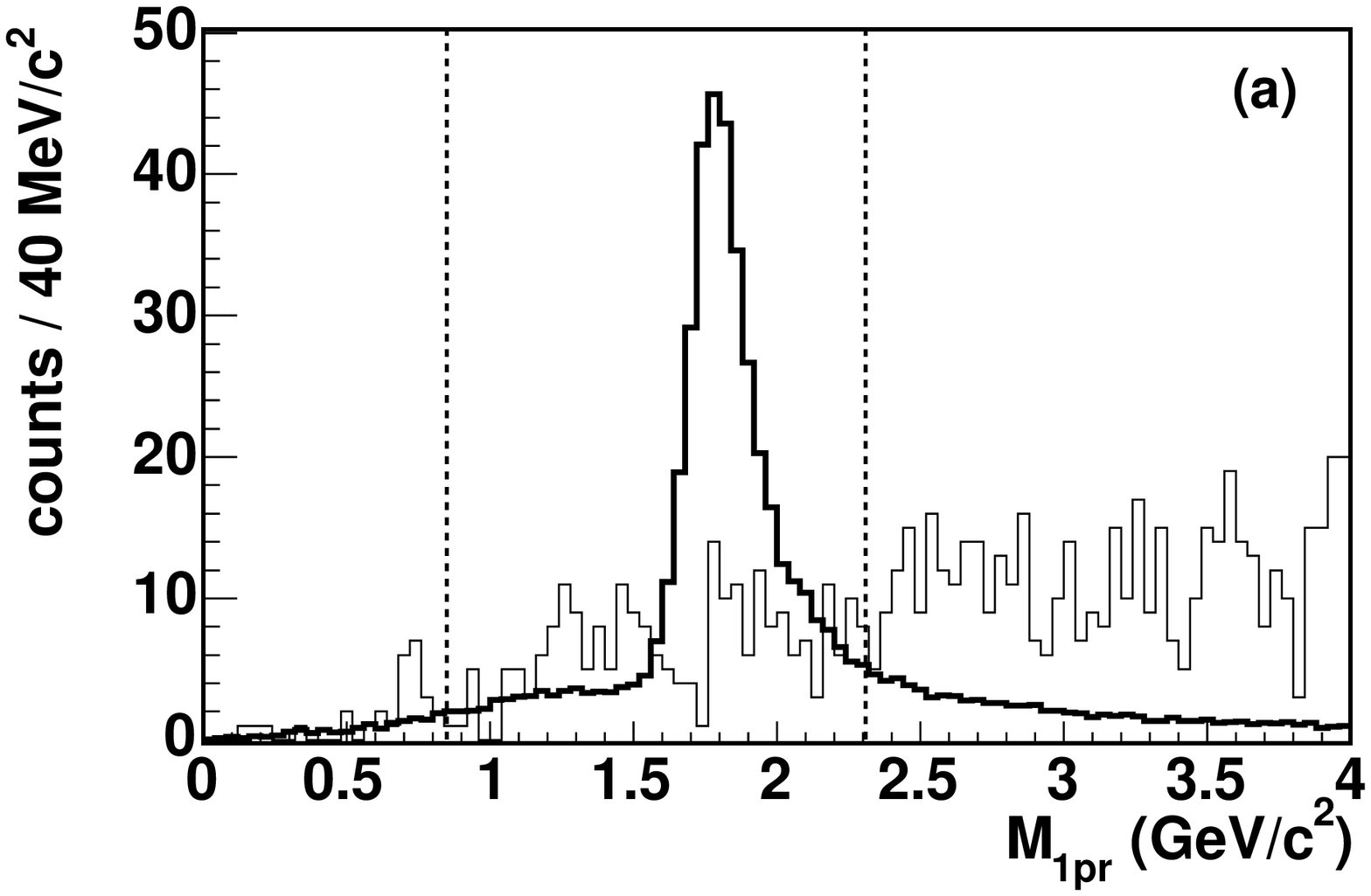}
\includegraphics[scale=0.43,clip]{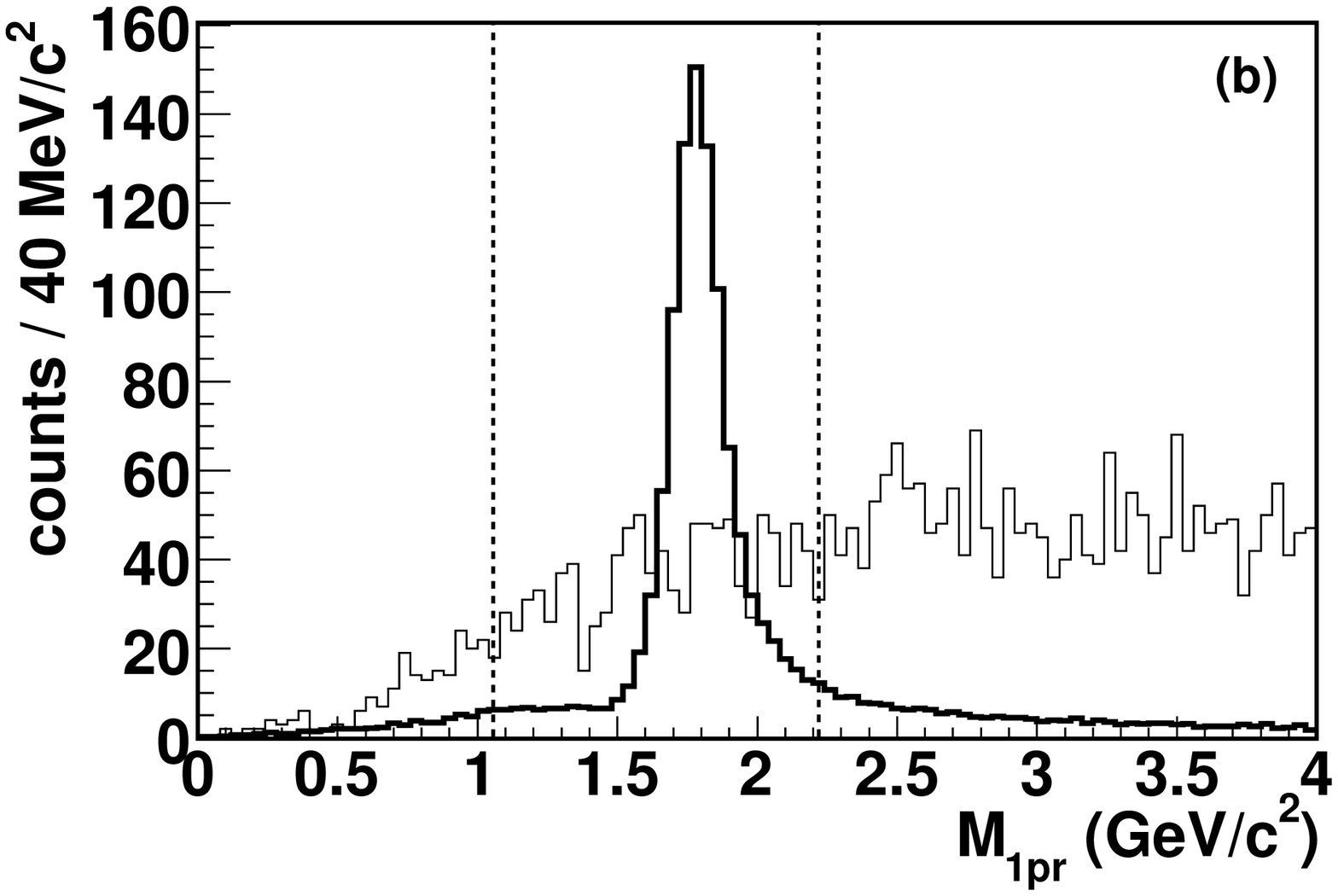}
\caption{$M_{\rm 1pr}$ distributions for the signal (thick line) and $uds$ continuum (thin line) MC events. The left figure (a) is for modes in which an electron is on the signal side, and the figure on the right (b) is for muon tags. Dashed lines show the boundaries of the signal region.}
\label{fig_m1p}
\end{center}
\end{figure}

We calculate the missing four-momentum for each event. 
In signal events, it is possible to reconstruct a $\tau$ mass from the missing momentum, 
1-prong charged track momentum and all momenta of photons on the tag side, since there is no 
neutrino emission from the signal decays. Figure \ref{fig_m1p} shows the distributions of the $\tau$ 
mass on the tag side $M_{\rm 1pr}$ for the signal events and $uds$ continuum background MC. 
There is a clear peak at the $\tau$ mass in the signal distribution, while the distribution for 
background process is smooth. 
Comparing the yield of these distributions, we optimize the
central value and width of the signal windows independently for electron and muon modes: 
0.85 GeV$/c^2 < M_{\rm 1pr} <$ 2.31 GeV$/c^2$ for the $\tau \to e h h$ and $\tau \to eV^0$ modes, and 
1.06 GeV$/c^2 < M_{\rm 1pr} <$ 2.22 GeV$/c^2$ for the $\tau \to \mu h h$ and $\tau \to \mu V^0$ modes.

To identify signal $\tau$ decays, we reconstruct the invariant mass $M_{\ell h h}$ 
and CMS energy $E^*_{\ell h h}$ of selected events. We then calculate the energy difference 
$\Delta E^* \equiv E^*_{\ell h h} - E^{*}_{\rm beam}$ and mass difference 
$\Delta M \equiv M_{\ell h h} - M_{\tau}$, where $E^{*}_{\rm beam}$ is the beam CMS energy 
and $M_\tau$ is the nominal $\tau$ mass. 
The signal events should have $\Delta E^*$ and $\Delta M$ around zero. 
We define the signal region in the $\Delta E^*$-$\Delta M$ plane using the signal MC. 
Signal MC distributions have tails on the lower sides of $\Delta E^*$ and $\Delta M$ 
due to effects from initial and final state radiation, bremsstrahlung of electrons and 
ECL energy leakage. 
We consider signal MC events as properly reconstructed if they satisfy the conditions 
$-0.68$ GeV $< \Delta E^* <$ 0.32 GeV and $-0.25$ GeV/$c^2 < \Delta M <$ 0.25 GeV/$c^2$.
We thus define a rectangular signal region that contains 90\% of properly reconstructed 
signal events. The boundaries of the signal region are summarized in Table \ref{tab:signalarea}.
\begin{table}
\caption{Definition of the signal regions for each decay mode.}
 \begin{tabular}{l c c}
  \hline
  Mode & ~~~~~$\Delta E^{*}$ (GeV) ~~~~~&  ~~~~~$\Delta M$  (GeV/$c^{2}$)~~~~~ \\
  \hline
  
$\tau^-\to e^-\pi^+\pi^-$   & $-0.10$ - +0.04  & $-0.014$ - +0.011\\  
  
$\tau^-\to e^+\pi^-\pi^-$   & $-0.09$ - +0.04  & $-0.014$ - +0.011\\  
  
$\tau^-\to \mu^-\pi^+\pi^-$ & $-0.07$ - +0.03  & $-0.011$ - +0.011\\  
  
$\tau^-\to \mu^+\pi^-\pi^-$ & $-0.07$ - +0.03  & $-0.011$ - +0.011\\  
  
$\tau^-\to e^-\pi^+K^-$     & $-0.10$ - +0.04  & $-0.013$ - +0.011\\  
  
$\tau^-\to e^-\pi^-K^+$     & $-0.10$ - +0.04  & $-0.012$ - +0.010\\  
  
$\tau^-\to e^+\pi^-K^-$     & $-0.10$ - +0.04  & $-0.014$ - +0.010\\  
  
$\tau^-\to e^-K^+K^-$       & $-0.10$ - +0.04  & $-0.010$ - +0.008\\  
  
$\tau^-\to e^+K^-K^-$       & $-0.10$ - +0.04  & $-0.013$ - +0.009\\  
  
$\tau^-\to \mu^-\pi^+K^-$   & $-0.08$ - +0.03  & $-0.009$ - +0.009\\  
  
$\tau^-\to \mu^-\pi^-K^+$   & $-0.07$ - +0.03  & $-0.009$ - +0.009\\  
  
$\tau^-\to \mu^+\pi^-K^-$   & $-0.08$ - +0.03  & $-0.009$ - +0.009\\  
  
$\tau^-\to \mu^-K^+K^-$     & $-0.07$ - +0.03  & $-0.007$ - +0.008\\  
  
$\tau^-\to \mu^+K^-K^-$     & $-0.07$ - +0.03  & $-0.007$ - +0.007\\  
  
$\tau^-\to e^- \rho^0$      & $-0.10$ - +0.04  & $-0.015$ - +0.012\\  
  
$\tau^-\to e^- K^*(892)^0$       & $-0.10$ - +0.04  & $-0.013$ - +0.011\\  
  
$\tau^-\to e^- \bar{K}^*(892)^0$ & $-0.08$ - +0.04  & $-0.012$ - +0.010\\  
  
$\tau^-\to e^- \phi$        & $-0.09$ - +0.03  & $-0.010$ - +0.008\\  
  
$\tau^-\to \mu^- \rho^0$    & $-0.07$ - +0.03  & $-0.011$ - +0.011\\  
  
$\tau^-\to \mu^- K^*(892)^0$       & $-0.08$ - +0.03  & $-0.009$ - +0.010\\  
  
$\tau^-\to \mu^- \bar{K}^*(892)^0$ & $-0.08$ - +0.03  & $-0.009$ - +0.009\\  
  
$\tau^-\to \mu^- \phi$      & $-0.08$ - +0.03  & $-0.007$ - +0.007\\  
  \hline
  \hline
 \end{tabular}
\label{tab:signalarea}
\end{table}

\section{Results}
The signal MC events are generated assuming a uniform decay angular 
distribution of $\tau$. Signal detection efficiencies $\epsilon$, evaluated from the 
MC, are listed in the fourth column of Table~\ref{tab:result} and vary from 2.68\% to 5.30\%.
The actual decay angle distribution, however, depends on the model for the LFV
interaction. 
In order to evaluate the possible effect 
of correlations, we examine $V-A$ and $V+A$ interactions using the 
formulae given in Ref. \cite{OKADA} and the relative differences in the efficiencies 
from the uniform distribution ($\Delta\epsilon/\epsilon$) are taken as systematic errors 
of detection efficiencies.

\begin{figure*}
\begin{center}
\includegraphics[scale=0.7,clip]{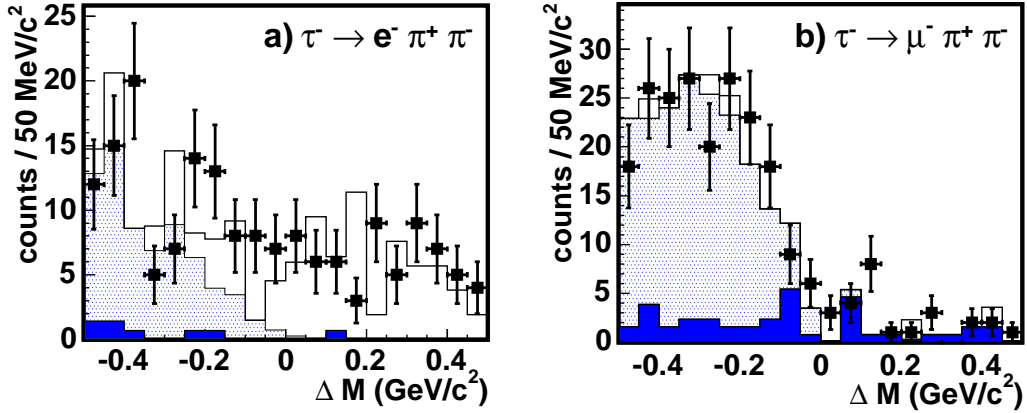}
\caption{$\Delta M$ distributions of experimental data (points with error bars) and expected background (histograms) for a) $\tau \to e^- \pi^+ \pi^-$ and b) $\tau \to \mu^- \pi^+ \pi^-$  modes after all selection criteria except for $\Delta M$. Particle identification corrections (see text) have been applied to all MC distributions. Different patterns of the background histograms correspond to various kinds of the background: $uds$ continuum (dark), $\tau$-pair generic decays (shaded) and two-photon process (blank). MC histograms are cumulative and show general agreement between the data and MC.} 
\label{Mbg}
\end{center} 
\end{figure*}

After all selection requirements, some events remain in the signal region. 
From MC, we evaluate the number of background events from $uds$ continuum 
and $\tau$-pair production. We then scale to the data by the factor 
$\sigma \mathcal{L}_{\rm int} / N_{\rm MC}$, where $\sigma$ is the cross section of this process, 
and $N_{\rm MC}$ is the number of generated events of the process. 
In the $\tau^- \to e^- h^+ h^-$ and $\tau^- \to e^- V^0$ modes, 
there is a contribution from two-photon $e^+e^- \to e^+e^-e^+e^-$, $e^+e^- \to e^+e^-\mu^+\mu^-$, 
$e^+e^- \to e^+e^-u\bar{u}$, $e^+e^- \to e^+e^-d\bar{d}$ and $e^+e^- \to e^+e^-s\bar{s}$ processes. 
Since the equivalent luminosity for the MC simulation of two-photon processes is much smaller than 
$\mathcal{L}_{\rm int}$, the contribution of the two-photon processes is estimated from a fit to 
the data in the $\Delta M$ sideband region, in which the contributions of $uds$ continuum and $\tau$-pair 
are fixed from MC and the two-photon shape is taken from MC with the normalization floated. 
For all modes the sideband region is  
$-0.5~{\rm GeV}/c^2 < \Delta M < 0.5~{\rm GeV}/c^2$, while the signal region 
$-96~{\rm MeV}/c^2 < \Delta M < 30~{\rm MeV}/c^2$ is blinded. The shape of the $\Delta M$ distribution 
is obtained after applying only the particle identification requirements. We verify that this shape 
does not change when additional selection criteria are applied by comparing the distribution 
before all selections and after each of them is successively imposed. 
It is known that $K/\pi$ separation is different for data and MC. 
We measure identification efficiency and fake rate for both of them 
using calibration samples obtained from $D^* \to D^0[K^-\pi^+] \pi$ decays 
and apply a correction to the MC distributions.  
The $\Delta M$ distributions of data and expected 
background for the $\tau^- \to e^- \pi^+\pi^-$ and $\tau^- \to \mu^- \pi^+ \pi^-$ modes after the 
corrections are shown in Figure \ref{Mbg}. The number of expected background events in the signal region 
for each studied decay mode is given in the fifth column of Table \ref{tab:result}. 
The uncertainty in the background expectation is dominated by MC statistics. 

\begin{figure*}
\begin{center}
\includegraphics[scale=0.5,clip]{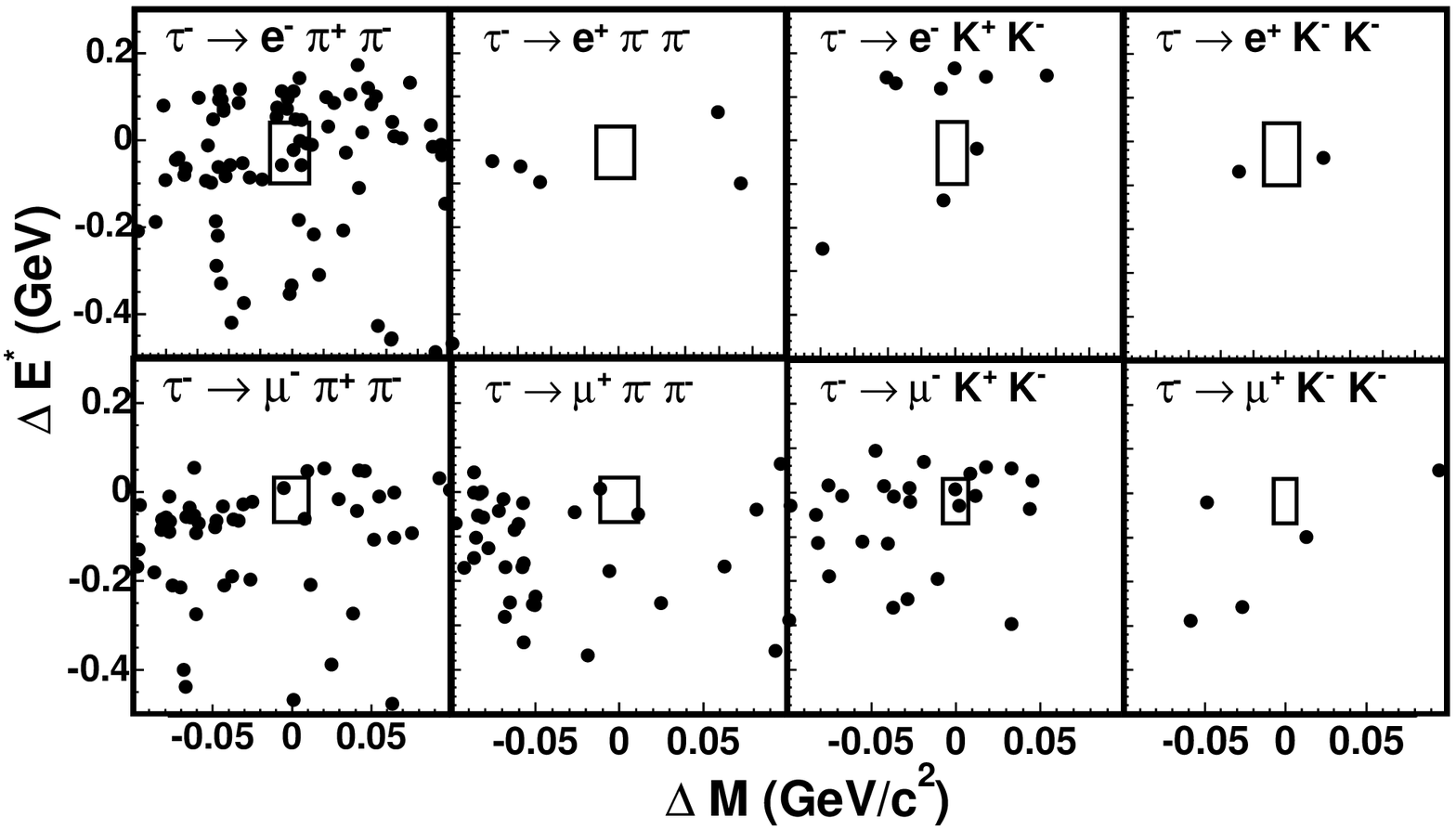}\\
\hspace*{0.cm}\includegraphics[scale=0.5,clip]{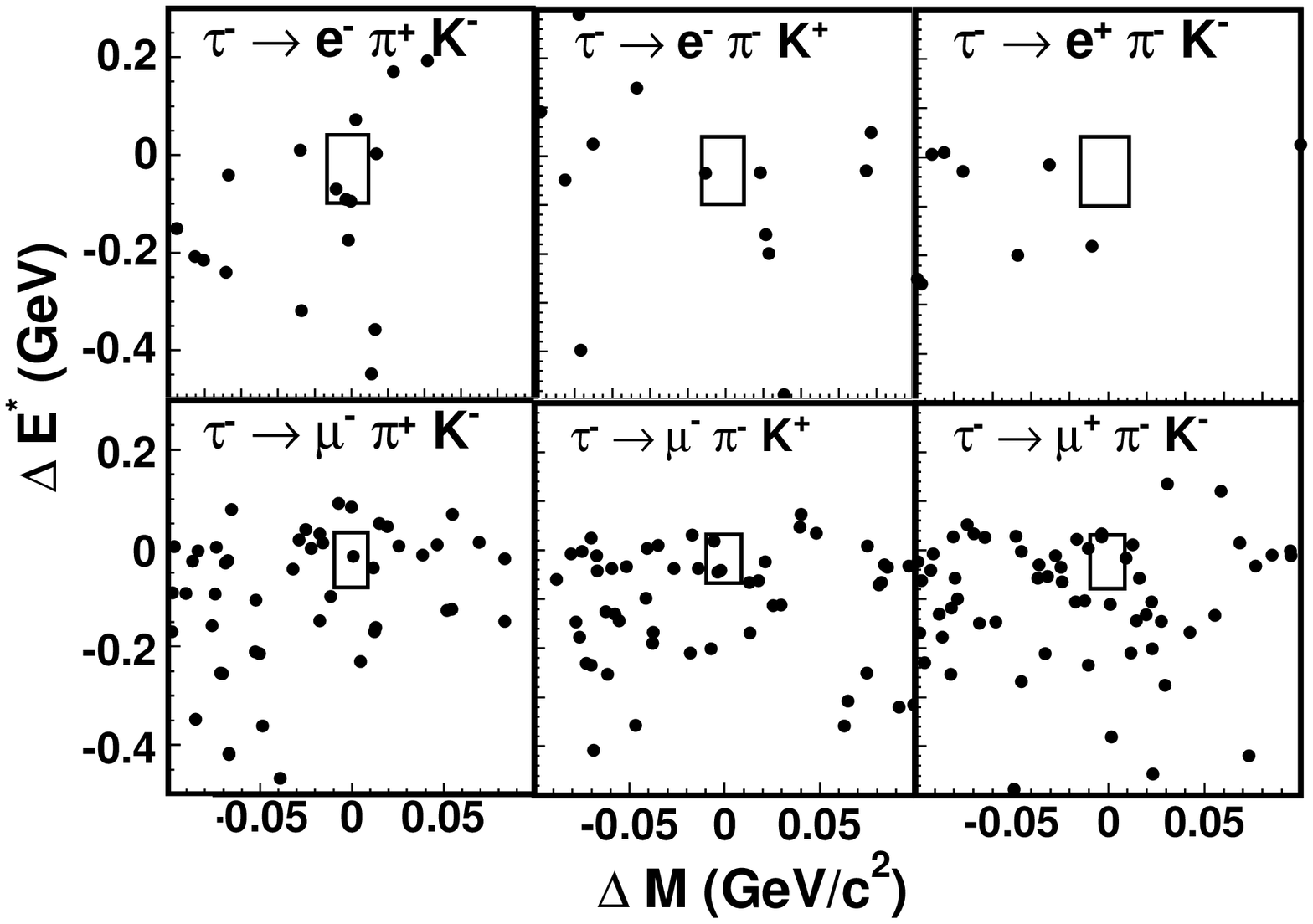}\\
\includegraphics[scale=0.5,clip]{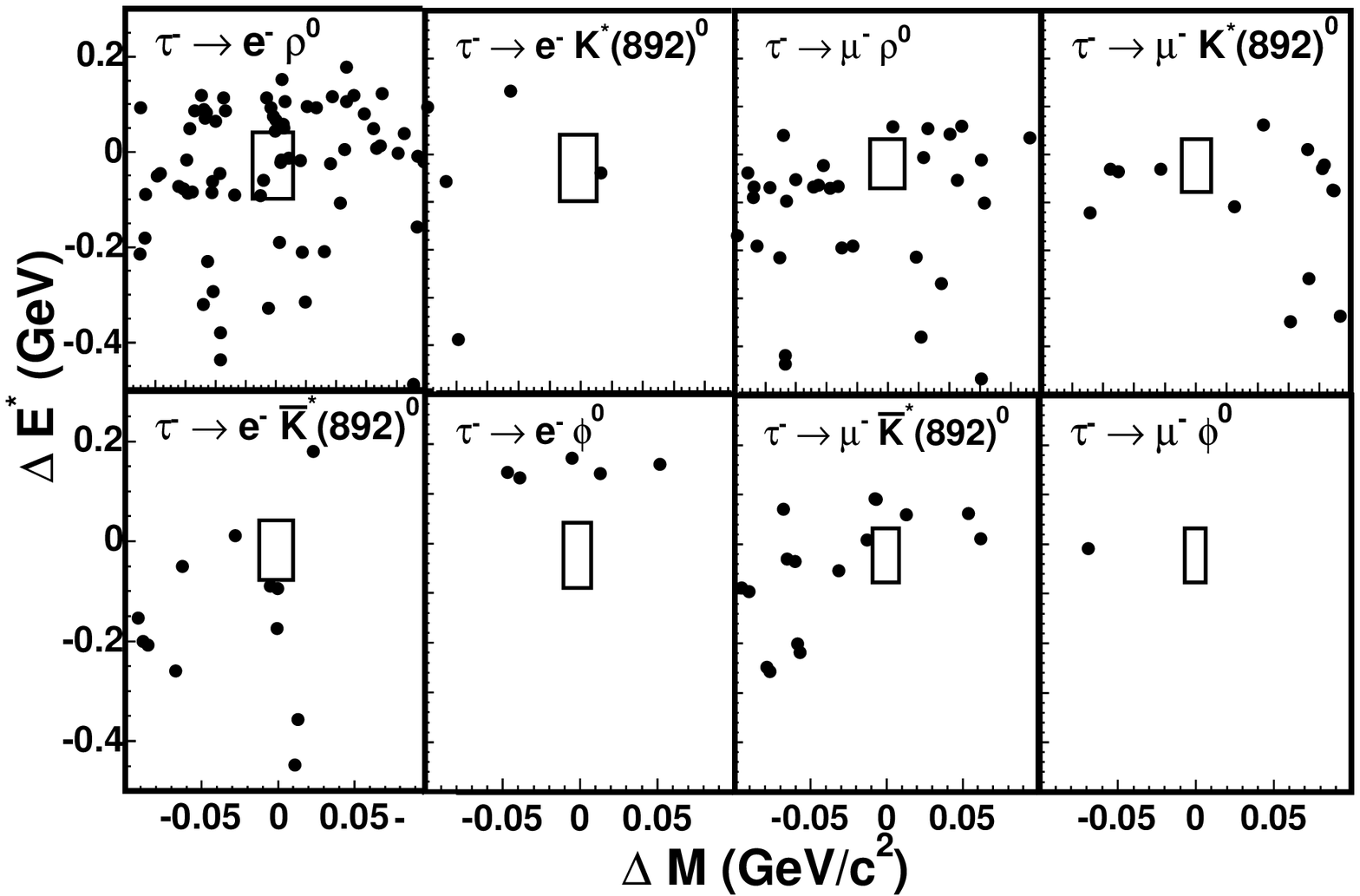} 
\caption{$\Delta E^{*}$ vs. $\Delta M$ experimental data distribution after all selection criteria. The boundaries of the signal region are illustrated by solid boxes.}
\label{EcmvsM}
\end{center}
\end{figure*}
\begin{table*}
\begin{center}
\caption{Summary of detection efficiency $\epsilon$ and its systematic error $\Delta{\epsilon}/\epsilon$, background expectation, number of observed events and 90\% C.L. upper limits on the branching fractions (BFs). For systematic uncertainty, LFV stands for the error due to the angular distribution of LFV decays, while ``Total'' combines LFV and all other errors.
}
\begin{tabular}{lcccccc}
  \hline 
Mode                           & \multicolumn{2}{c}{$\Delta{\epsilon}/\epsilon$~(\%)}  & \multicolumn{1}{l}{Detection}                  & \multicolumn{1}{l}{Expected} & \multicolumn{1}{l}{Observed}  & \multicolumn{1}{l}{Upper limit on} \\
                               &LFV &Total                                       & \multicolumn{1}{l}{efficiency $\epsilon$ (\%)} & \multicolumn{1}{l}{background}  & \multicolumn{1}{l}{events}    & \multicolumn{1}{l}{BF (90\% CL)}\\
  \hline

$\tau^- \to e^- \pi^+ \pi^-$   & 5.3  & 7.5  & 5.30 & 2.62$\pm$1.07  & 6 & 7.3$\times 10^{-7}$\\

$\tau^- \to e^+ \pi^- \pi^-$   & 2.3  & 5.8  & 5.14 & 0.00$\pm$0.26  & 1 & 2.0$\times 10^{-7}$\\

$\tau^- \to \mu^- \pi^+ \pi^-$ & 2.1  & 8.8  & 4.37 & 0.76$\pm$0.26  & 2 & 4.8$\times 10^{-7}$\\

$\tau^- \to \mu^+ \pi^- \pi^-$ & 7.7  & 11.5 & 4.44 & 0.73$\pm$0.30  & 1 & 3.4$\times 10^{-7}$\\

$\tau^- \to e^- \pi^+ K^-$     & 20.5 & 21.2 & 3.99 & 0.91$\pm$0.25  & 3 & 7.2$\times 10^{-7}$\\

$\tau^- \to e^- \pi^- K^+$     & 17.4 & 18.2 & 4.11 & 1.27$\pm$0.41  & 0 & 1.6$\times 10^{-7}$\\

$\tau^- \to e^+ \pi^- K^-$     & 12.8 & 13.9 & 4.03 & 0.74$\pm$0.22  & 0 & 1.9$\times 10^{-7}$\\

$\tau^- \to e^- K^- K^+$       & 21.9 & 22.5 & 3.12 & 0.34$\pm$0.20  & 0 & 3.0$\times 10^{-7}$\\

$\tau^- \to e^+ K^- K^-$       & 5.4  & 7.6  & 3.06 & 0.09$\pm$0.07  & 0 & 3.1$\times 10^{-7}$\\

$\tau^- \to \mu^- \pi^+ K^-$   & 15.8 & 18.0 & 3.43 & 2.35$\pm$0.44  & 1 & 2.7$\times 10^{-7}$\\

$\tau^- \to \mu^- \pi^- K^+$   & 19.1 & 20.9 & 3.32 & 1.85$\pm$0.32  & 3 & 7.3$\times 10^{-7}$\\

$\tau^- \to \mu^+ \pi^- K^-$   & 25.4 & 26.8 & 3.53 & 2.53$\pm$0.38  & 1 & 2.9$\times 10^{-7}$\\

$\tau^- \to \mu^- K^- K^+$     & 8.7  & 12.2 & 2.76 & 0.48$\pm$0.19  & 2 & 8.0$\times 10^{-7}$\\

$\tau^- \to \mu^+ K^- K^-$     & 38.2 & 39.2 & 2.70 & 0.09$\pm$0.06  & 0 & 4.4$\times 10^{-7}$\\

$\tau^- \to e^- \rho^0$             & 5.3  & 7.5  & 5.03 & 2.55$\pm$1.04  & 5 & 6.5$\times 10^{-7}$\\

$\tau^- \to e^- K^*(892)^0$         & 17.4 & 18.2 & 4.12 & 0.76$\pm$0.34  & 0 & 3.0$\times 10^{-7}$\\

$\tau^- \to e^- \bar{K}^*(892)^0$   & 20.5 & 21.2 & 3.68 & 0.16$\pm$0.10  & 0 & 4.0$\times 10^{-7}$\\

$\tau^- \to e^- \phi$               & 21.9 & 22.5 & 2.94 & 0.04$\pm$0.04  & 0 & 7.3$\times 10^{-7}$\\

$\tau^- \to \mu^- \rho^0$           & 2.1  & 8.8  & 4.40 & 0.26$\pm$0.12  & 0 & 2.0$\times 10^{-7}$\\

$\tau^- \to \mu^- K^*(892)^0$       & 19.1 & 20.9 & 3.61 & 0.37$\pm$0.14  & 0 & 3.9$\times 10^{-7}$\\

$\tau^- \to \mu^- \bar{K}^*(892)^0$ & 15.8 & 18.0 & 3.42 & 0.49$\pm$0.19  & 0 & 4.0$\times 10^{-7}$\\

$\tau^- \to \mu^- \phi$             & 8.7  & 12.2 & 2.68 & 0.00$\pm$0.18  & 0 & 7.7$\times 10^{-7}$\\
  \hline
  \hline
 \end{tabular}
\label{tab:result}
\end{center}
\end{table*}
The $\Delta E^*$ vs. $\Delta M$ plots for the experimental data for all decay modes 
are shown in Figure \ref{EcmvsM}. The numbers of events observed in the signal regions are 
listed in the sixth column of Table \ref{tab:result}. They are consistent with those 
expected from background distributions.
We set the upper limits $s_0$ on the number of the signal events at 90\% CL 
using the prescription of Feldman and Cousins~\cite{Poisson-limit}. 
The main systematic uncertainties on the detection efficiency come 
from track reconstruction (1.0\% per track), electron identification 
(1.1\% per electron), muon identification (5.4\% per muon), kaon/pion 
separation (1.0\% per kaon and pion), trigger efficiency (1.4\%), statistics of the signal MC (1.0\%)
and uncertainties of the branching fractions of vector meson decays (1.2\% for $\phi \to K^+ K^-$). 
The uncertainty on the number of $\tau$-pair events mainly comes from the luminosity measurement (1.4\%). 
The systematic uncertainties due to the angular distribution of LFV $\tau$ decays are summarized in 
the second column of Table \ref{tab:result}, together with the total systematic errors 
$\Delta\epsilon/\epsilon$, which are used to evaluate the total errors of the sensitivities.
To evaluate $s_0$, we use the Poisson probability density function, 
which is convolved with the uncertainties on the sensitivities and expected background 
assuming the Gaussian shape \cite{Cousins-Highland,POLE}.

Upper limits on the branching fractions $\mathcal{B}$ are calculated 
as $\mathcal{B}(\tau \to \ell h h) < \displaystyle 
\frac{s_0}{2 N_{\tau\tau}  \epsilon  \mathcal{B}_1}$ and 
$\mathcal{B}(\tau^{-} \to \ell V^0) < \displaystyle \frac{s_0}{2 N_{\tau\tau}  \epsilon \mathcal{B}_{1}  \mathcal{B}_{\rm V}}$, where $N_{\tau\tau}$ is the total number of the $\tau$-pairs produced, 
$\mathcal{B}_1$ is the inclusive branching fraction for the 1-prong decay of $\tau$ 
and $\mathcal{B}_{\rm V}$ is the branching fraction of vector meson decay 
into charged hadrons. We use $\mathcal{B}_1 = (85.35 \pm 0.07)\%$ and
$\mathcal{B}_\phi = (49.2 \pm 0.6)\%$ from the 2005 web update of Ref. \cite{PDG}. 
The resulting upper limits on the branching fractions are  summarized in 
the last column of Table~\ref{tab:result}. 
\section{Discussion}
Our final results for 90\% upper limits on the branching fractions for the 
$\tau \to \ell h h$ modes, shown in Table \ref{tab:result}, are in the range 
$(1.6 - 8.0) \times 10^{-7}$.
Our results for $\tau \to \ell h h$ and $\tau \to \ell V^0$ modes are one order of magnitude more 
restrictive than those obtained in the CLEO experiment~\cite{CLEOLHH} except for a few decay modes. 
The results for $\tau^- \to \ell h h$ modes are comparable to those from BaBar~\cite{BABARLHH}. 

While in many extensions of the Standard Model the predicted values of
the branching fractions of LFV decays are very small and out of reach
for current experiments, some models allow enhancements of such decays to a 
level very close to the experimentally accessible range at the $B$ factories.
For example, detailed analysis of various LFV $\tau$ lepton decays in
the framework of MSSM show that at small $\tan{\beta}$ and 
appropriate values  of other model parameters the branching ratio of 
the $\tau^- \to \mu^- \rho^0$ decay can be as high as $10^{-8}$~\cite{bb}. 

The improved sensitivity to rare $\tau$ lepton decays 
achieved in this work can be used to constrain the parameters of 
models with heavy Dirac neutrinos~\cite{seesawilakovac00}. In this model the 
expected branching fractions of various LFV decays are evaluated  
in terms of combinations of the model parameters. These combinations, 
denoted $y^2_{\tau e}$ and $y^2_{\tau \mu}$ for $\tau$ decays involving 
an electron and a muon, respectively, can vary from 0 to 1. 
Our best 90\% upper limit for the modes with an electron, 
$y^2_{\tau e} < 0.24$, can be set from the $\tau^- \to e^- \rho^0$ decay. 
The best corresponding limit for modes with a muon studied in this 
work is set from the $\tau^- \to \mu^- \rho^0$ mode, 
$y^2_{\tau \mu} < 0.38$. This bound is more restrictive than 
any other limits set on LFV decays of the $\tau$.

Our results can also be used to constrain energy scale of new physics 
in models with dimension-six effective fermionic operators that 
induce $\tau - \mu$ mixing~\cite{black}. From our upper limits for 
the branching fractions of $\tau^- \to \mu^-\rho^0$, $\tau^- \to \mu^-\phi$ and 
$\tau^- \to \mu^- K^{*}(892)^0$ the bounds $\Lambda_{uu,dd} >29.4$~TeV, 
$\Lambda_{ss} > 24.8$~TeV and  $\Lambda_{ds} > 26.8$~TeV, respectively, can 
be obtained for the models with vector operators. 
\section{Summary}
We have searched for LFV decays $\tau \to \ell h h$ and $\tau \to \ell V^0$ using a 158.0 fb$^{-1}$ 
($140.9 \times 10^6$ $\tau$-pair events) data sample in the Belle experiment. 
No evidence for a signal of these decay 
modes is observed and upper limits on the branching fractions are set in the range 
$(1.6-8.0) \times 10^{-7}$, which are one order of magnitude more restrictive than 
those previously obtained by the CLEO experiment. For $\tau \to  \ell h h$ modes, the 
results are comparable to recent limits from BaBar \cite{BABARLHH}.
\vspace*{1cm}\\
We thank the KEKB group for the excellent operation of the
accelerator, the KEK cryogenics group for the efficient
operation of the solenoid, and the KEK computer group and
the National Institute of Informatics for valuable computing
and Super-SINET network support. We acknowledge support from
the Ministry of Education, Culture, Sports, Science, and
Technology of Japan and the Japan Society for the Promotion
of Science; the Australian Research Council and the
Australian Department of Education, Science and Training;
the National Science Foundation of China and the Knowledge Innovation Program of Chinese Academy of Sciencies under contract No.~10575109 and IHEP-U-503; the Department of Science and Technology of
India; the BK21 program of the Ministry of Education of
Korea, and the CHEP SRC program and Basic Research program 
(grant No. R01-2005-000-10089-0) of the Korea Science and
Engineering Foundation; the Polish State Committee for
Scientific Research under contract No.~2P03B 01324; the
Ministry of Science and Technology of the Russian
Federation; the Slovenian Research Agency;  the Swiss National Science Foundation; the National Science Council and 
the Ministry of Education of Taiwan; and the U.S. 
Department of Energy.

\end{document}